\documentclass[aps, prd, showpacs, superscriptaddress, twocolumn, nofootinbib, preprintnumbers, groupedaddress]{revtex4}
\usepackage[utf8]{inputenc}
\usepackage{bm}
\usepackage{latexsym,amssymb,amsmath,amsfonts}
\usepackage{graphicx}
\graphicspath{{fig/}}
\usepackage[usenames]{color}
\usepackage[breaklinks, backref, colorlinks=true]{hyperref}

\def\d{\textrm{d}}
\def\GW{\textrm{GW}}
\def\OGW{\Omega_\text{GW}}
\def\rhoc{\rho_\textrm{c}}

\begin{document}

\preprint{IUCAA-27/2015}

\title{Stochastic Gravitational Wave Background from Exoplanets}

\author{Anirban~Ain}
\email{ainz@iucaa.ernet.in}
\affiliation{Inter-University Centre for Astronomy and Astrophysics, Post Bag 4, Ganeshkhind, Pune 411007, India}

\author{Shilpa~Kastha}
\email{shilpakastha@imsc.res.in}
\affiliation{Department of Physics, Visva-Bharati University, Santiniketan 731235, India}
\altaffiliation[Current Affiliation: ]{The Institute of Mathematical Sciences, Chennai 600113, India}

\author{Sanjit~Mitra}
\email{sanjit@iucaa.ernet.in}
\affiliation{Inter-University Centre for Astronomy and Astrophysics, Post Bag 4, Ganeshkhind, Pune 411007, India}

\begin{abstract}


Recent exoplanet surveys have predicted a very large population of planetary systems in our galaxy, more than one planet per star on the average, perhaps totalling about two hundred billion. These surveys, based on electro-magnetic observations, are limited to a very small neighbourhood of the solar system and the estimations rely on the observations of only a few thousand planets. On the other hand, orbital motions of planets around stars are expected to emit gravitational waves (GW), which could provide information about the planets not accessible to electro-magnetic astronomy. The cumulative effect of the planets, with periods ranging from few hours to several years, is expected to create a stochastic GW background (SGWB). We compute the characteristic GW strain of this background based on the observed distribution of planet parameters. We also show that the integrated extragalactic background is comparable or  less than the galactic background at different frequencies. Our estimate shows that the net background is significantly below the sensitivities of the proposed GW experiments in different frequency bands. However, we notice that the peak of the spectrum, at around $10^{-5}$~Hz, is not too far below the proposed space based GW missions. A future space based mission may be able to observe or tightly constrain this signal, which will possibly be the only way to probe the galactic population of exoplanets as a whole.



\end{abstract}
  
\maketitle

\section{Introduction}
\label{intro}

Einstein's theory of General Relativity predicts that massive bodies with oscillating mass-quadrupole moment emit Gravitational Waves (GW)~\cite{gw,MTW}. Planets orbiting around stars are therefore expected to emit GW. While the signal from a single planet may be too weak to detect by the current and upcoming GW detectors~\cite{Berti1,Berti2}, it was important to check if the huge number of planets predicted by the recent exoplanet missions, like Kepler~\cite{KeplerMission}, can constitute a detectable stochastic GW background (SGWB). SGWB is an incoherent superposition of signals from a large set of sources where an individual source can not be resolved~\cite{AllenSchool}. SGWBs of different cosmological and astrophysical origin have been estimated in literature~\cite{AllenSchool,Grishchuk00,Turner96,isoLVC_2009-10,BinCap12,Marassi11,CowTan06, Zhu12, Rosado11, RaviWyi12, Mingarelli13, hotspot,Mazumder2014,BenderHils,HilsBenderWebbink,WDSGWB} and several upper limits have been placed using data from various experiments and many sophisticated analysis techniques~\cite{AllenSchool,BBNBound,SmithEtAl,RottiSouradeep,WMAP9par,PlanckParam,sgwbS5iso,isoLVC_2009-10,sgwbS5dir,Jenet2006,Haasteren2011,NanoGravSGWB,PPTASGWB}. In this paper we estimate the background arising from the predicted population of exoplanets in the Milky Way galaxy due to their orbital motion in their planetary system. The current exoplanet missions have observed few thousand planets in a small neighbourhood of the solar system. Based on these measurements a population model was estimated, predicting more than one planet per star, totalling about two hundred billion planets in our galaxy~\cite{OneOrMorePlanets}. Using the observed distributions of planet parameters (mass, ellipticity, orbital period and semi-major axis) from the publicly available exoplanet databases we estimate the frequency spectrum of the SGWB created by all the planetary systems in our galaxy. We also compute the strength of the background from the Andromaeda galaxy and check if the extragalactic component, integrated over the rest of the universe, has any significant contribution as compared to its galactic counterpart.

We then explore the possibility of detection of this background. Though direct detection of GW has not been possible yet, observation of decay in orbital period of binary pulsar PSR B1913+16 over three decades provides a convincing evidence of the existence of GW~\cite{HulseTaylor,WeisbergNiceTaylor}. Direct detection of GW is one of the most important challenges in the current research in Astrophysics and Cosmology, which is likely to happen in the next few years, perhaps through different windows of observation. While the ground based laser interferometric GW observatories~\cite{AdvLIGO,AdvVirgo,KAGRA} are most promising for detecting GW sources at $\sim 100$Hz, a low frequency stochastic background may be detected soon though the measurements of the Cosmic Microwave Background ``B-mode'' polarisation anisotropy~\cite{BICEP2,PlanckPolDust, BICEP2-Planck} and the Pulsar Timing Arrays (PTA)~\cite{IPTA,EPTA,NanoGrav,PPTA}. The proposed space based detectors~\cite{eLISA, DECIGO, BBO}, planned to be launched in the next decades, are expected to observe different GW signals at very high Signal-to-Noise Ratio (SNR) compared to the other detectors. For this work PTA and space based detectors are the relevant ones, as their frequency bands have intersections with the frequency range of the planet orbits. We discuss the possibility of detection of the SGWB from exoplanets using these two kinds of detectors.


The paper is organised as follows: we estimate the SGWB created by galactic and extragalactic exoplanets in Section~\ref{Estimation}. Possibility of its detection is considered in Section~\ref{Detection}. We conclude with discussions in Section~\ref{Discussions}.


\section{Estimation of SGWB}
\label{Estimation}

\subsection{Formalism}
\label{formalism}

The waveform and flux of GW from two point masses in a Keplerian orbit has been calculated very precisely~\cite{RadiationKepler}. If a two-body system of masses $M$ and $m$ are moving in Keplerian orbits around the centre of mass with effective semi-major axis $a$, the total GW power radiated by the system (a.k.a. GW Luminosity), averaged over one period of the elliptic motion, is
\begin{equation}
L_0 \ = \ \frac{32}{5}\frac{G^4}{c^5}\frac{M^2m^2(M+m)}{a^5} \;.
\end{equation}
The average frequency of this wave is $f_{0} =\omega_{0} / \pi$, where
\begin{equation}
\omega_{0}=\sqrt{{G(M+m)}/{a^3}} \, .
\label{eq:Freq}
\end{equation}
If the system has an eccentricity $e$ then the total power emitted increases and becomes
\begin{equation}
L_0 \ = \ \frac{32}{5}\frac{G^4}{c^5}\frac{M^2m^2(M+m)}{a^5{(1-e^2)}^\frac{7}{2}}\left(1+\frac{73}{24}e^2 + \frac{37}{96}e^4\right) \;.
\end{equation}
In the eccentric case the radiation is no longer monochromatic. The total power radiated in the $n^\mathrm{th}$ harmonic, at a frequency $f=n \omega_0 / \pi$, is given by
\begin{equation}
L_n \ = \ \frac{32}{5}\frac{G^4}{c^5}\frac{M^2m^2(M+m)}{a^5}g(n,e) \, ,
\label{eq:Power}
\end{equation}
where,
\begin{equation}
\begin{split}
g(n,e) &:= \frac{n^4}{32}\bigg(\big[ J_{n-2}(ne)-2eJ_{n-1}(ne) \ + \\
\frac{2}{n} J_n(ne) & + 2eJ_{n+1}(ne)-J_{n+2}(ne)\big]^2 \ + (1-e^2) \times \\
[J_{n-2}(ne) & - 2J_n(ne)+J_{n+2}(ne)]^2 +\frac{4}{3n^2}{[J_n(ne)]}^2\bigg) \, .
\label{eq:harmonics}
\end{split}
\end{equation}
The energy density received on earth from a source of luminosity $L$ is $\rho = L/4 \pi d^2 c$. We express the energy spectrum in terms of the usual dimensionless quantity $\OGW(f)$ defined as the energy density per unit logarithmic frequency interval in the units of the average critical density required for a spatially flat universe, $\rho_c = 3 H_0^2 c^2/ 8\pi G$, where $H_0 := 100 \, h_{100}$~km/s/Mpc is the Hubble constant at the current epoch,
\begin{equation}
\OGW(f) \ := \ \frac{1}{\rhoc} \frac{\d \rho_\GW}{\d \ln f} \, .
\end{equation}
The above quantity can be converted to characteristic GW strain~\cite{AllenSchool,Jenet2006} through the equation
\begin{equation}
h_{\rm c}(f)  \ = \ 1.5 \times 10^{-18} \, \sqrt{h_{100}^2 \OGW (f)} \, f^{-1} \, , \label{eq:Strain}
\end{equation}
which is easier to compare with experimental sensitivity.

\subsection{Back of the Envelope Calculation}
\label{envelope}

Before indulging into the detailed numerical computation of the background, we first obtain a back of the envelope estimate for a simple system with parameters reasonably close to the actual observed values presented in the next section. Consider a uniformly dense spherical galaxy of radius $R \sim 17$kpc with $N \sim 2 \times 10^{11}$ stars all of the same mass $M = 1 M_\odot$ and each one with a planet of Jupiter mass $m = 10^{-3}~M_\odot$ in circular orbit with a period of $10^7 - 10^8$~sec. If the planets are distributed uniformly in logarithmic frequency interval, the average number density of the planets per unit frequency is given by $N / f \ln 10$.  Since the GW frequency is twice the orbital frequency, the total GW energy density per unit logarithmic frequency interval received at the centre of this galaxy is given by
\begin{equation*}
\frac{\d \rho_\GW}{\d \ln f} = \frac{f}{c} \int _0^R \frac{L(f)}{4 \pi r^2} \frac{N}{{4 \over 3} \pi R^3 f \ln 10} \, 4 \pi r^2 \d r = \frac{3 N L(f)}{4 \pi R^2 c \ln 10} \, ,
\end{equation*}
where $L(f)$ is the average GW luminosity of one planet with wave frequency $f$ (i.e., orbital frequency $f/2$). For extreme mass ratio binaries such as these, where $M \gg m$, $L(f) \approx (32 G^{7/3}/ 5 c^5) M^{4/3} m^2 (\pi f)^{10/3}$ [see Eq.~(\ref{eq:Freq})].
Putting all together one gets
\begin{eqnarray}
\OGW (f) &=&\frac{24 \, G^{7/3} N M^{4 / 3} m^2}{(5 \pi \ln{10}) \, c^6 \rhoc R^2}  (\pi f)^{10/3} \\
&\approx& 1.4 \times 10^{-26} \, h_{100}^{-2} \, \left( \frac{f}{10^{-8} \rm{Hz}} \right)^{10/3} \, ,
\end{eqnarray}
and in terms of characteristic GW strain
\begin{equation}
h_{\rm c}(f) \ = \ 1 \times 10^{-23} \, \left( \frac{f}{10^{-8} \rm{Hz}} \right)^{2/3} \, .
\end{equation}
If instead of a spherical galaxy, we take a disk of the same radius and of thickness $600$~pc, the estimate doubles
\begin{equation}
h_{\rm c}(f) \ = \ 2 \times 10^{-23} \, \left( \frac{f}{10^{-8} \rm{Hz}} \right)^{2/3} \, .
\end{equation}
This $h_{\rm c}(f)$ is overlaid on Fig~\ref{fig:spectra} with a dashed line along with more detailed numerical estimation presented in the next section. Though simplistic, this calculation, matches the numerical estimate, including the power law index, reasonably well.

Few more comments are in order here. The numbers used in the back of the envelope calculation are motivated from the distributions obtained from real data and the spectrum shifts by orders of magnitude if the numbers are not reasonably precise. So, strictly speaking, this is not a back the envelope calculation. The reason for choosing this frequency range of $10^{-8} - 10^{-7}$~Hz for this calculation was that the histogram of the orbital periods (not shown in the paper) peaks in this range, so the estimation error will be minimal. Although the background seems low in this range, it has a rising spectrum and only full numerical evaluation could give a full spectrum and hence analyse the detectability of the background.
 
\subsection{Radiation from confirmed exoplanets}
\label{sec:measured}

  In the last few decades an extraordinary number of exoplanets has been discovered. The number of known exoplanets is growing rapidly. We collected the orbital data of known exoplanets from the Exoplanet Orbit Database~\cite{EOD}. Among the $1499$ confirmed planets in the database, the parameters required to calculate GW emissions are available for $596$ planets. Most of these planets are within $100$ parsecs from earth. We assume the parameters semi-major axis and eccentricity do not change significantly over the observing period.
  
  \begin{figure}
  \includegraphics[width=0.4\textwidth,height=0.2\textheight]{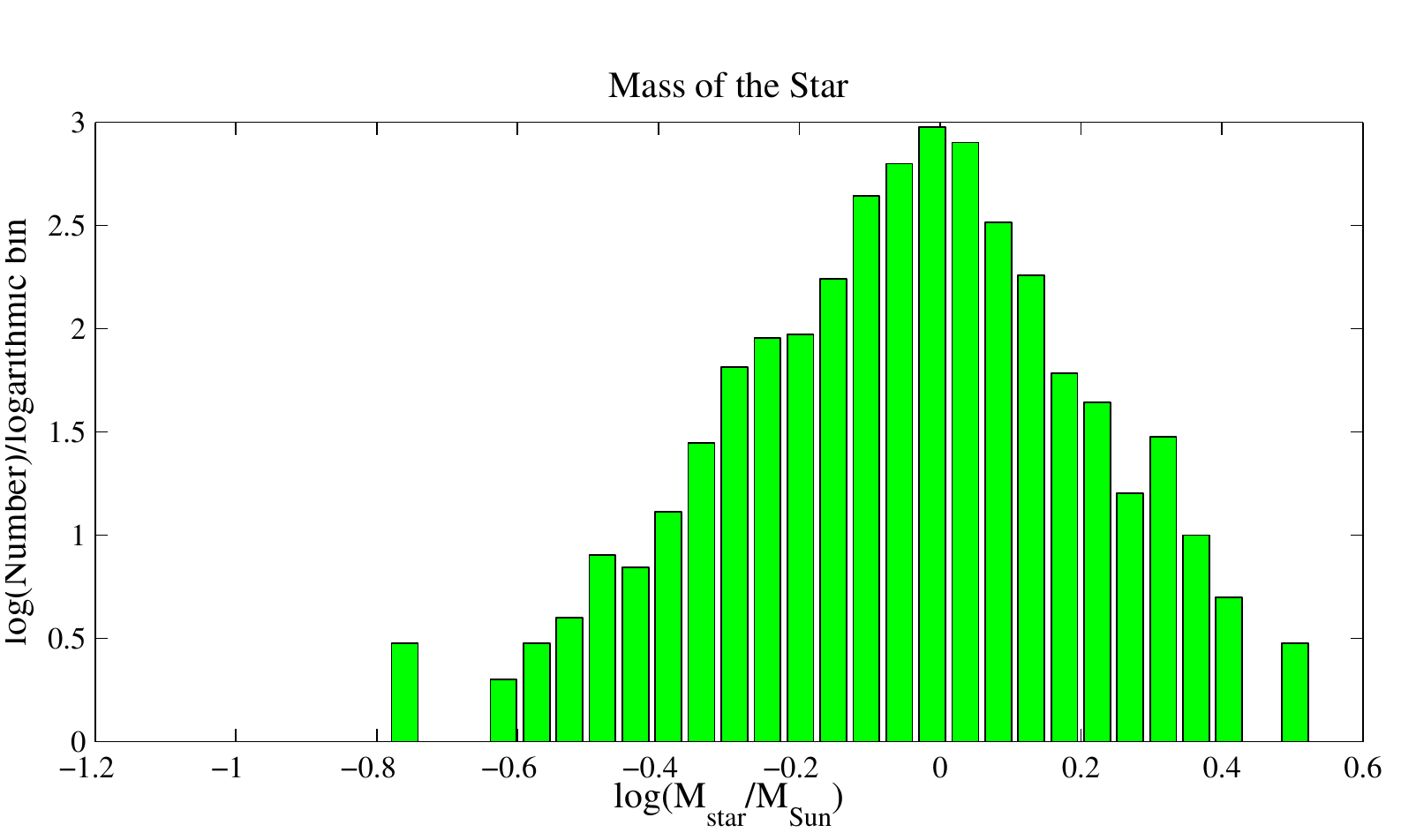}
  \includegraphics[width=0.4\textwidth,height=0.18\textheight]{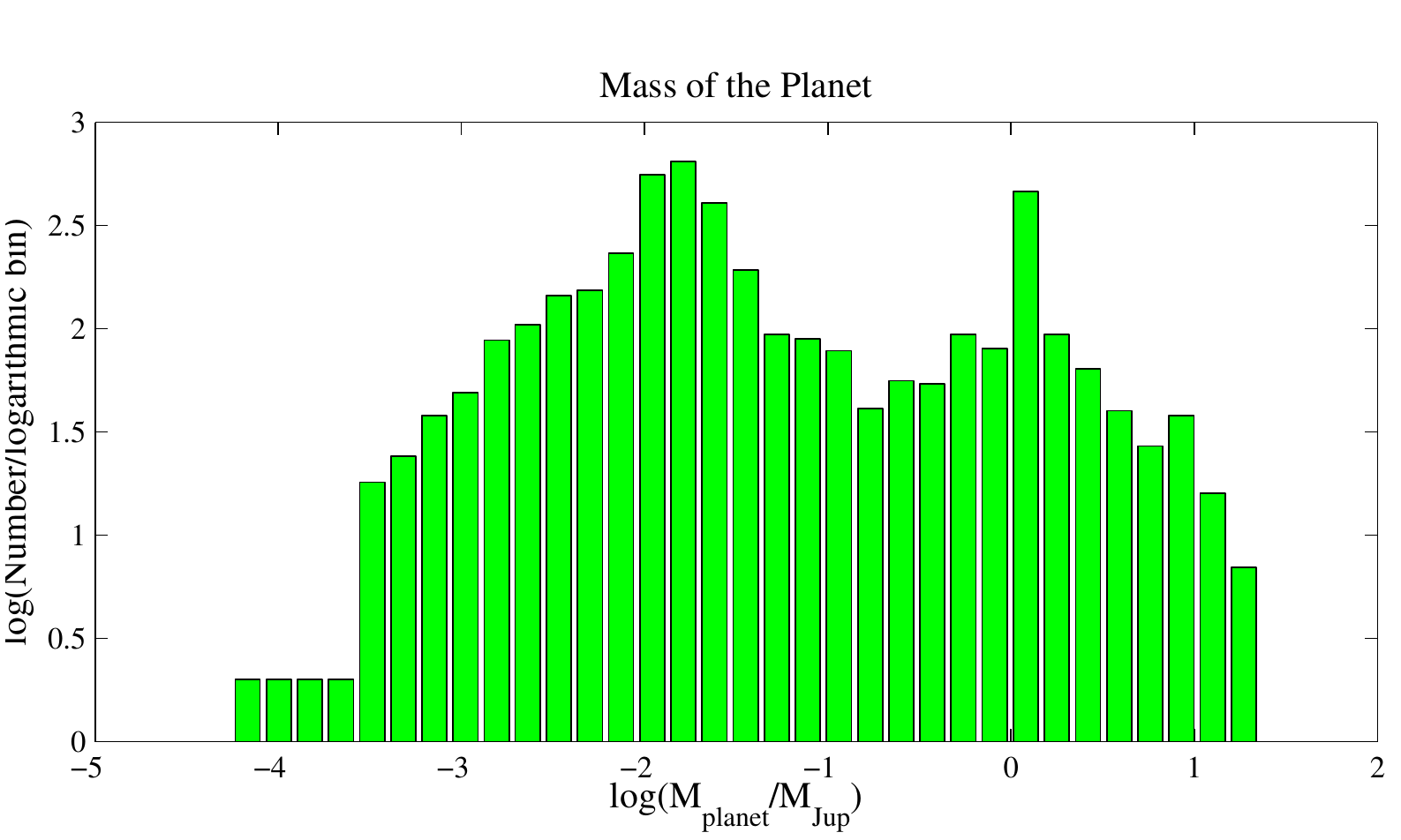}
  \includegraphics[width=0.4\textwidth,height=0.18\textheight]{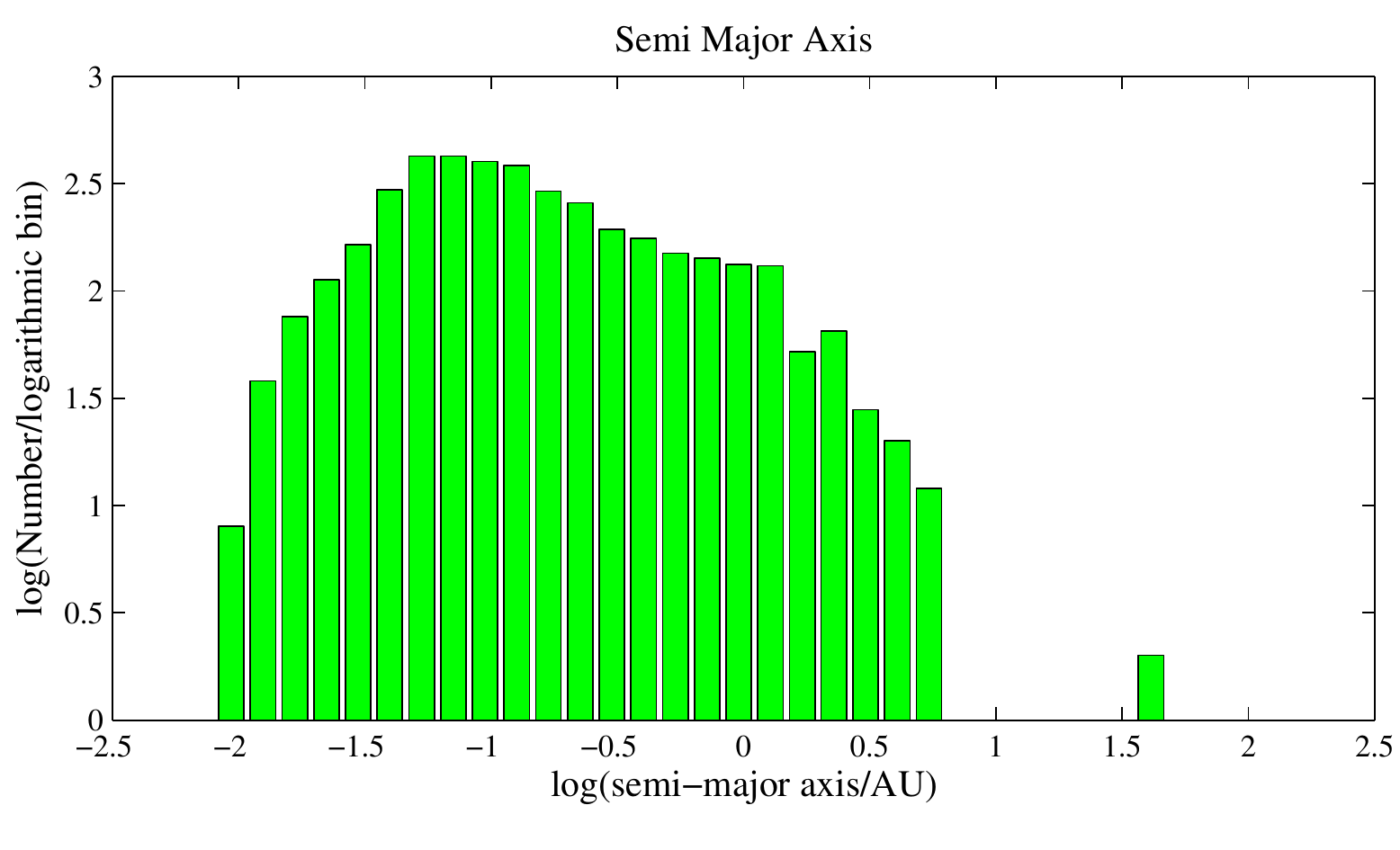}
  \includegraphics[width=0.4\textwidth,height=0.18\textheight]{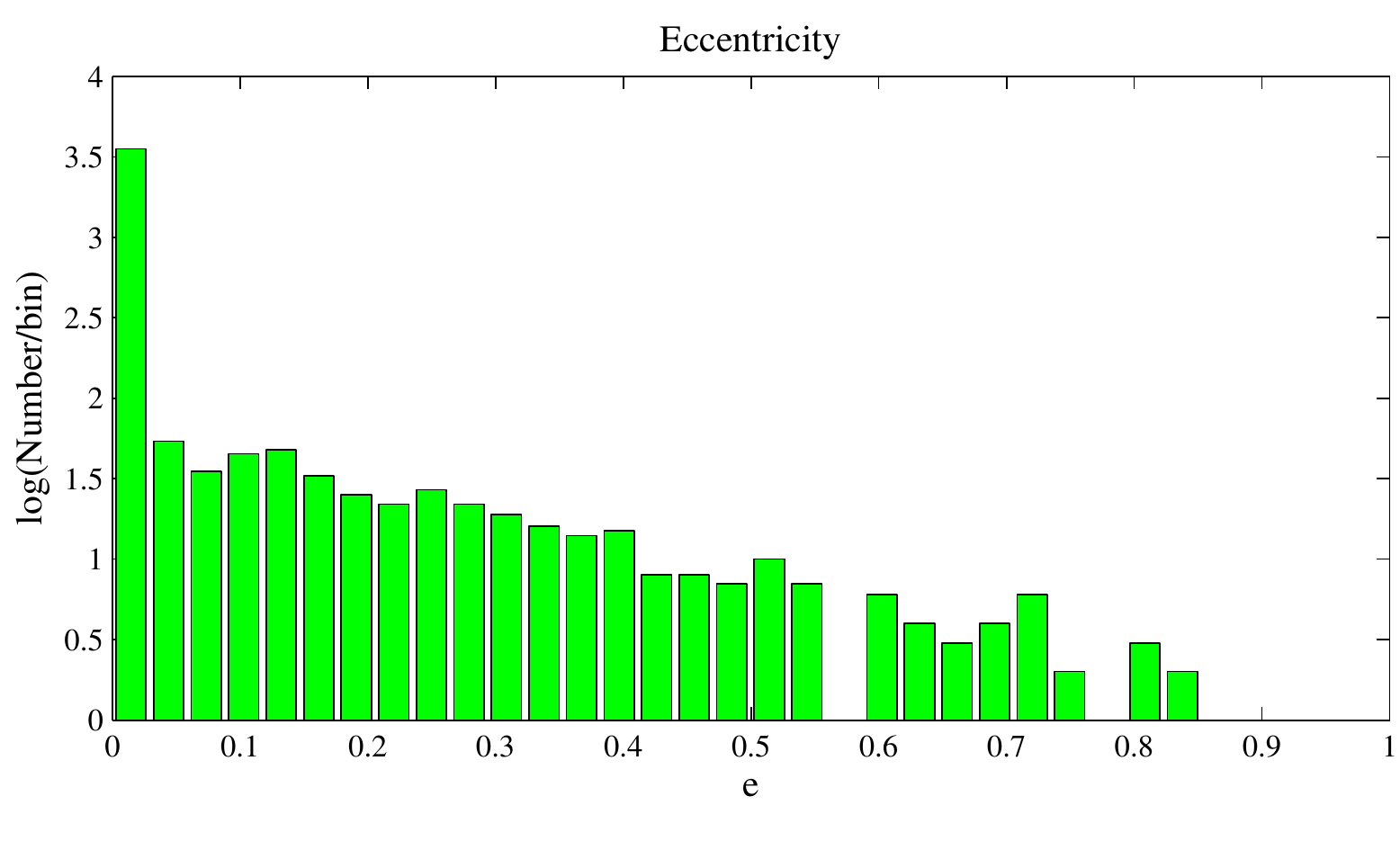}
  \includegraphics[width=0.4\textwidth,height=0.18\textheight]{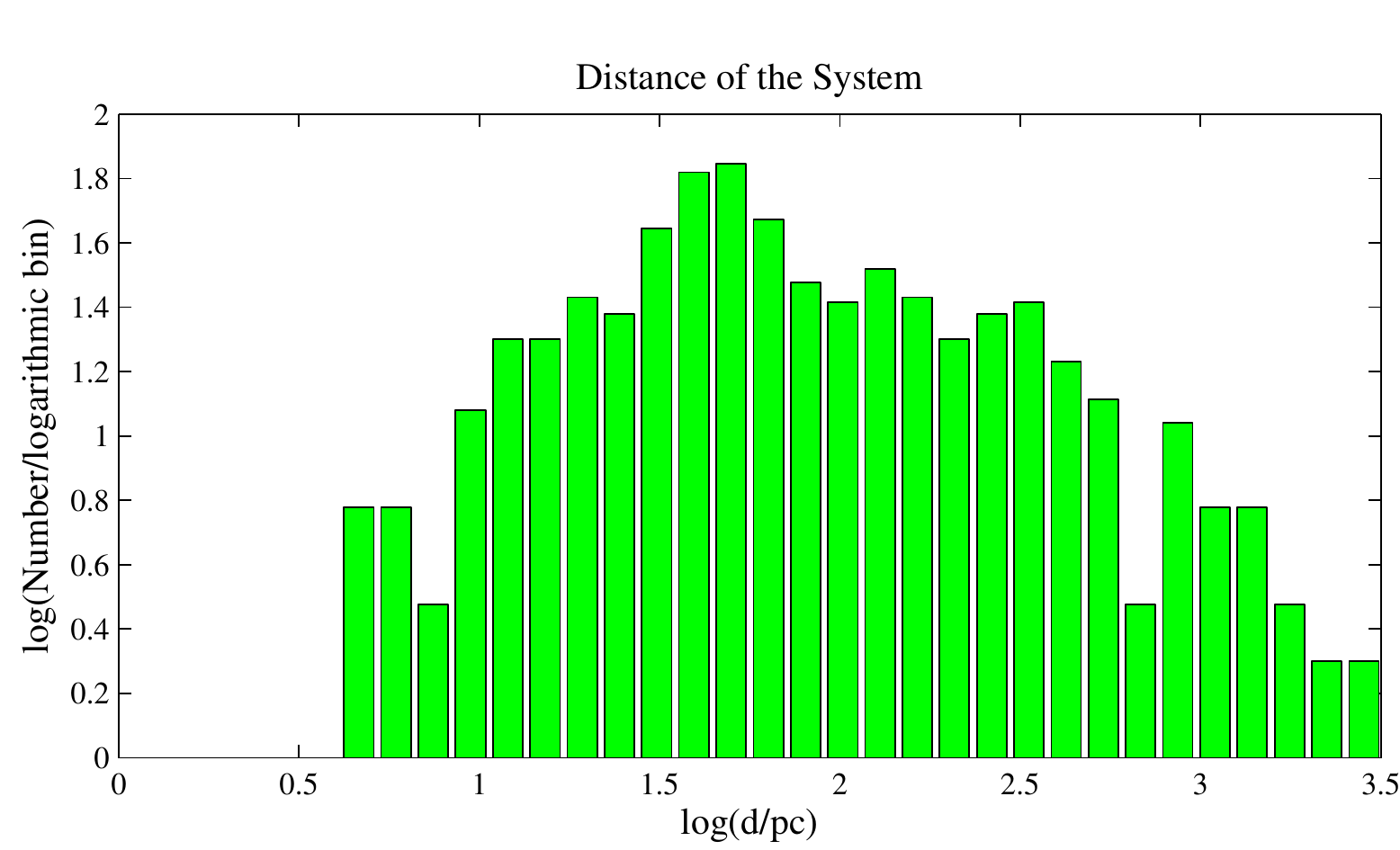}
  \caption{\label{fig:parhist} We make histograms of measured planet parameters obtained from the exoplanet databases. The bottom histogram has been created from $596$ observed for which the distances were known, out of the total $3988$ planets, which were used for the top four histograms. Wherever eccentricity was not measured, it was set to zero to get a more conservative estimate. The simulations are done by drawing samples from the distributions of masses, semi-major axis and eccentricity.}
  \end{figure}

\begin{figure*}
\centering
\includegraphics[width=0.9\textwidth]{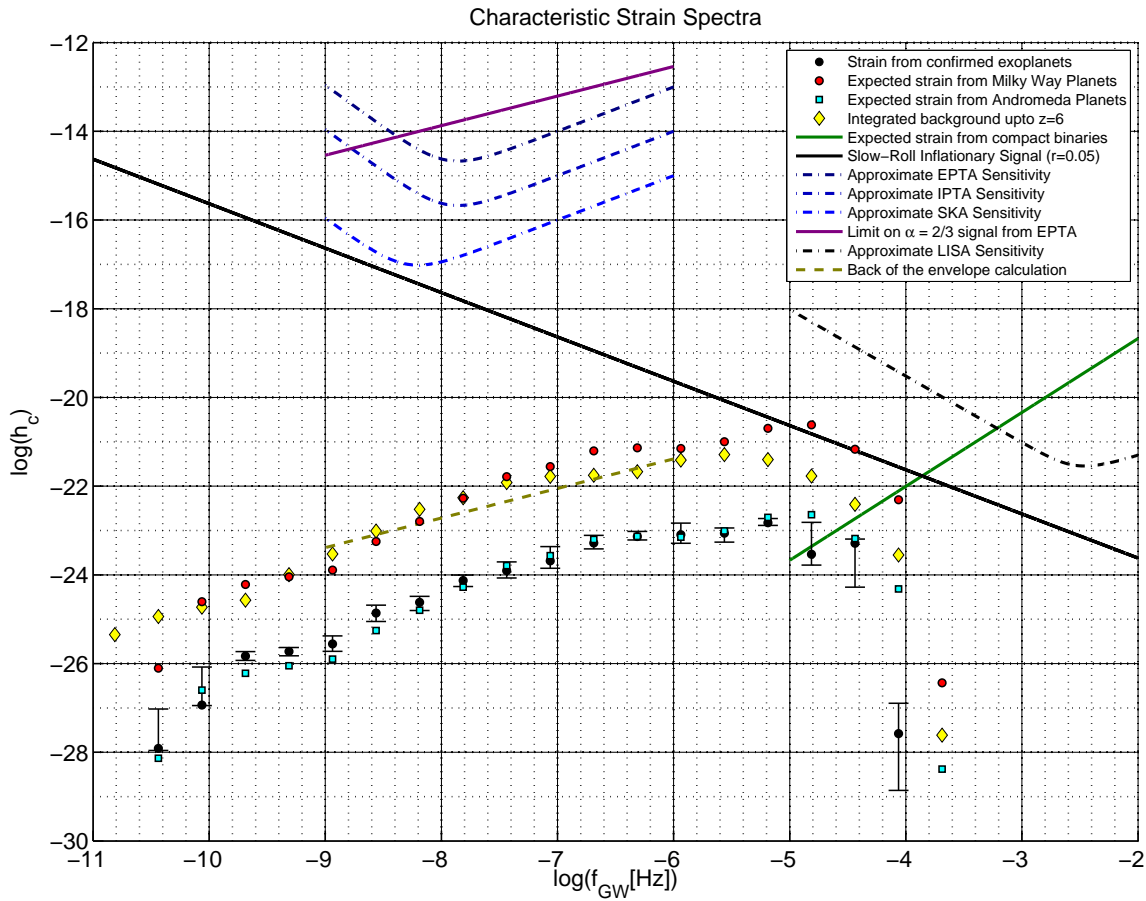}
\caption{\label{fig:spectra} Our estimated characteristic strains and sensitivities and upper limits of LISA and different PTAs are shown here. The black circles with error bars provide the background from the observed planets. When the observed distribution of parameters is used for simulating all the planets in the galaxy, we get the background from the whole galaxy denoted by solid red dots, which is the main result of this paper. In frequencies higher than $10^{-5}$ the background is almost equal to the Inflationary Background~\cite{InflationSignal} for a tensor-to-scalar ratio of $r=0.05$ (overlaid in solid black line). In that frequency range the signal is higher than that of compact binaries in our galaxy~\citep{SGWBCompactBinaries} plotted in solid green line). The sensitivities of different PTAs~\cite{PTASensitivity,GWSensGraph} (plotted in bluish dot-dashed curves) are way above the strength of planetary GW. The $95$\% upper-limit line from EPTA data~\cite{Haasteren2011}, corresponding to a spectral index of $\alpha = +2/3$, is plotted with a purple solid curve. The LISA sensitivity is in black dot-dashed curve. The total characteristic strain for all the planets in the Andromeda galaxy is also overlaid in cyan squares. The  total background created by all other galaxies in the universe up to $z=6$ in LCDM cosmology and uniform comoving galaxy density is shown with yellow diamonds.}
\end{figure*}

We then calculate the power radiated by each planetary system and the frequencies in which the power is radiated using Eq.~(\ref{eq:Freq}-\ref{eq:Power}). The total spectrum is calculated by adding the average flux received from all the planets. Though GW radiation is not isotropic from a single source, while averaging over a large number of sources with no preferred orientation, it is reasonable to assume isotropic emission. The characteristic strain spectrum, obtained from eqn.~(\ref{eq:Strain}) is shown in Fig.~\ref{fig:spectra} by black circular dots with errorbars.

\subsection{SGWB from galactic exoplanets}

The main aim of this paper is to estimate the background from all the planets of the Milky Way galaxy, which we do via simulations. Studies of exoplanets suggest that there might be one or more planets per star in our galaxy \cite{OneOrMorePlanets}. We first compute the distributions of parameters of the detected exoplanets, taking into account the errorbars, as shown in Fig.~\ref{fig:parhist}. We do not observe any noticeable correlation between the parameters. We estimate the background by adding flux from sample star-planet systems with parameters randomly drawn from the neighbourhood of their observed values  and randomly placed with uniform distribution at different parts of the galaxy. We assume that there are $2 \times 10^{11}$ stars each with only $1$ planet, though the actual number can be more. In reality, some stars have many planets and some have none, but since in general each planet emits at a distinct frequency, it is statistically equivalent to distribute the planets uniformly among all the stars. Here we have taken the shape of the galaxy to be a circular disk of radius $17$~kpc and of thickness $0.6$~kpc and placed the solar system $8.34$~kpc away from the centre of the disk, where we estimate the background. Since the ``quadrupole formula'' used for calculating the Luminosity breaks down and flux appears to diverge if the distance is zero, in order to avoid random samples to appear too close to the solar system, we exclude the samples closer than $1000$pc from the solar system. We fill this ``hole'' with samples from actual observations, which lie mostly within $1000$pc, as seen in Fig.~\ref{fig:parhist}. The backgrounds estimated from this simulations are shown in Fig.~\ref{fig:spectra} by red circular points. We discuss the detectability of this background in the next section.

Exoplanet detection methods have different selection biases, which are in turn sensitive to different combinations of masses and orbit size. The confirmed exoplanets are more likely to have low orbit size and high masses and this region of the parameter space is also more significant for gravitational wave emission. Thus the estimate provided above should be close to reality. Also the distribution of mass of stars should be described by an Initial Mass Function (IMF). So we redid the simulation with the star mass distributed as an IMF~\cite{IMF} and the results are almost identical to the previous case.


\subsection{SGWB from extragalactic exoplanets}

We also estimate the background created by the Andromaeda galaxy (cyan squares in Fig.~\ref{fig:spectra}) $r_\text{A} \sim 800$kpc away from us, assuming the same parameter distributions as the Milky Way, except for distance from the earth, which is practically the same for all the planets in Andromaeda, and its mass, which is $\sim 2$ times that of the Milky Way galaxy. The background is $\sim 1$\% of the Milky Way background strain. If one substitutes a constant distance $r = r_\text{A}$ in our back of the envelope calculation, one can show that the background characteristic strain is $\sim \sqrt{2/6} \, R / r_\text{A} \sim 1$\% times that from the Milky Way, consistent with the numerical result. Similarly, since the Virgo cluster is $\sim 1500$ times more massive than the Milky Way and about $r_\text{V} = 16.5$~Mpc away, the characteristic strain from Virgo is $\sim \sqrt{1500/6} \, R / r_\text{V} = 1.6$\% of the Milky Way, slightly greater than that from the Andromaeda galaxy.

Finally we estimate the background for all the planets in the universe by integrating over different redshifts. The spectrum of background from different galaxies falls off with distance and gets redshifted, but the number of galaxies increase with distance. The combined effect requires numerical evaluation for different cosmological models. We apply a procedure similar to that of~\citet{Mazumder2014} for the similar case of exoplanets. If each galaxy emitted $J(f) \d f$ amount of energy in the frequency range $f$ to $f + \d f$ per unit time, per unit Milky Way Equivalent Galaxy (MWEG) mass, the integrated background spectra is given by
\begin{equation}
\OGW(f) \ = \ \frac{\pi}{3 c^2 H_0^3} f \int_0^\infty \d z \, \frac{8 G \, J(f(1+z)) \, n(z) }{a^2(t_0) (1+z) \, E(z)} \, ,
\end{equation}
where, $n(z)$ is the comoving number density of MWEGs at a given redshift $z$, $a(t_0)$ is the cosmological scale factor at the present epoch $t_0$ (generally scaled to $1$) and $E(z)$ is the Hubble parameter, which can be expressed, in terms of fractional matter and dark energy densities $\Omega_\Lambda$ and $\Omega_M$, as $E(z) = \sqrt{\Omega_\Lambda + (1+z)^3 \, \Omega_M}$ for a spatially flat universe.
Assuming that the comoving Milky Way equivalent galaxy density of the universe is $n(z) \sim 0.01 h_{100}^2 \rm{Mpc}^{-3}$~\cite{longairGalaxy}, a constant, and integrating up to  a cosmological redshift of $z \sim 6$ with a statistically isotropic $\Lambda$CDM model, we find that in this particular case the integrated extragalactic background characteristic strain [shown in Figure~\ref{fig:spectra} by yellow diamonds] is lower than the Milky Way background at the higher frequencies and comparable or little higher at the lower frequencies. It is worth noting that the extragalactic background is expected to be isotropic, while the galactic background is expected to be mostly limited in the galactic plane, hence the latter should stand out in the extragalactic background.
 
\section{Possibilities for Detection}
\label{Detection}

\subsection{Pulsar Timing Array}
Pulsar are very precise clocks. However the presence of a stochastic background can add ``jitter'' to the pulse timing. Studying the variance of this jitter for a set of well-studied pulsars, a ``Pulsar Timing Array'' (PTA)~\cite{Lorimer-rev}, can provide an upper limit to the stochastic background. The correlations between these jitters from different pulsars not only improve the upper limit, but can also provide directional information for localised sources on the sky.  PTA is particularly sensitive in the frequency range of $10^{-9}$ to $10^{-7}$ Hz. The expected sensitivities of different PTAs, in terms of characteristic strain, are overlaid in Figure~\ref{fig:spectra} with dot-dashed curves of different shades of blue~\cite{PTASensitivity,GWSensGraph}.

Our predicted background is clearly not detectable with current International PTA (IPTA)~\cite{IPTA} and does not seem plausible even with SKA-PTA. We overlay the EPTA limit on SGWB from \citet{Haasteren2011}, which provided upper-limits for different values of the spectral index $\alpha$. Since a power law of the form $f^{2/3}$ fits the planetary background reasonable well, we read off the $95$\% upper limit on characteristic strain corresponding to $\alpha=2/3$ and overlay with a solid purple line.

\subsection{LISA and eLISA}
The Laser Interferometer Space Antenna(LISA) and its incarnation Evolved Laser Interferometer Space Antenna (eLISA) are triplets of satellites arranged in an equilateral triangle with million kilometre arms using laser interferometry to measure GW. The sensitivity of these systems go as low as $10^{-5}$~Hz~\cite{LISASensitivity} The expected sensitivity LISA mission is overlaid in Fig~\ref{fig:spectra} with black dash-dot curve~\cite{LISASensitivity}.

The predicted background is closer to the LISA sensitivity than any other detection method. Though the background peaks at this frequency range, it also drops drastically in this band. It could be due to some selection bias in planet detection. Nevertheless, the number of planets among the Kepler Candidates with orbital frequency above $10^{-5}$~Hz is not small, so the statistical error in that frequency band is not very large and the estimate is conservative.

\section{Discussions}
\label{Discussions}

In this paper we estimate the Stochastic Gravitational Wave Background (SGWB) spectrum produced by all the  planetary systems in our galaxy. Even when there are nearly two billion such planetary systems in our galaxy the total background is still small.
%
%
We however see a significant amount of background in the band of the proposed space based detector LISA ($\sim 10^{-5}$~Hz). Though the total strain is still couple of orders of magnitude below LISA sensitivity, one can hope that a future space based detector will achieve such a sensitivity.

There are planets detected which have orbital period of few days and also there are lower frequency planets with higher eccentricity which could have harmonics in this band. It is intriguing to note that the planetary background almost smoothly merges with the galactic compact binary background, the later is far more dominant at $\gtrsim 10^{-4}$~Hz~\cite{SGWBCompactBinaries}.

Generally the SGWB considered in literature are due to extra-galactic sources and the analysis is done assuming plane waveform. In case of galactic sources, however, one may need special care to account for spherical wavefronts. The exoplanets of Milky Way lie in the galactic plane, hence one may also have to account for the shape of the galaxy to improve detection SNR by performing a search for an anisotropic background~\cite{TaylorGairAniso}.

Electro-magnetic astronomy has allowed observations of only a very small fraction of galactic exoplanets, much less than a millionth! Which certainly limits the average statistical information on planets obtained form those observations. Accounting for the selection bias in detection, for instance, would require an independent handle for better understanding of the population distribution of the planets. In summary, if the SGWB from the planets is detected, we believe it will  shine new light on our understanding of average statistical properties of galactic planet population.
 
\begin{acknowledgments}
We would like to thank Varun Bhalerao, Sukanta Bose, Sanjeev Dhurandhar, Bhooshan Gadre, Charles Jose, Aditya Rotti, Tarun Souradeep and the Stochastic Group of the LIGO-Virgo collaboration for useful discussions. This research has made use of the Exoplanet Orbit Database and the Exoplanet Data Explorer at exoplanets.org~\cite{EOD}. AA acknowledges the support of Council of Scientific and Industrial Research (CSIR), India. SM acknowledges the support of Science and Engineering Research Board (SERB), India for the FastTrack grant SR/FTP/PS-030/2012.
\end{acknowledgments}
  
\bibliography{planetSGWB}

\end{document}